\documentclass{PoS}
\newcommand{\be}{\begin{equation}} \newcommand{\ee}{\end{equation}}
\newcommand{\ba}{\begin{eqnarray}} \newcommand{\ea}{\end{eqnarray}}
\newcommand{\bea}{\begin{eqnarray}} \newcommand{\eea}{\end{eqnarray}}
\newcommand{\bean}{\begin{eqnarray*}} \newcommand{\eean}{\end{eqnarray*}}

\newcommand{\s}[1]{{\scriptscriptstyle #1}}
\newcommand{\st}{{\s T}}

\title{Time-reversal-odd phenomena in QCD}

\ShortTitle{T-odd phenomena in QCD}

\author{\speaker{P.J. Mulders}\\
        Department of Physics and Astronomy, FEW, VU University\\
        De Boelelaan 1081, NL-1009 DB Amsterdam, the Netherlands\\
        E-mail: \email{mulders@few.vu.nl}}

\abstract{
Single-spin asymmetries were long thought to vanish in high-energy
scattering processes because of their specific time-reversal behavior.
Time-reversal-odd phenomena, however, appear naturally when one 
includes effects of intrinsic transverse momenta of partons. 
The partons, quarks and gluons, enter the description of 
high-energy scattering processes in correlators which contain quark and gluon 
fields. The correlators, parameterized 
in terms of distribution and fragmentation functions, constitute
matrix elements of non-local field configurations.
For transverse momentum dependent (TMD) correlators the non-locality is
along a light-front, in contrast to the non-locality along the light-cone
for collinear correlators, integrated over transverse momenta. 
The TMD correlators require a careful treatment to assure color 
gauge invariance, leading to nontrivial gauge links connecting 
the parton fields. These give rise to time-reversal-odd phenomena,
showing up as single spin and azimuthal asymmetries.
The gauge links, arising from multi-gluon initial and final state
interactions, depend on the color flow in the process, which has 
consequences for issues like universality and factorization.
}

\FullConference{LIGHT CONE 2008 Relativistic Nuclear and Particle Physics\\
		 July 7-11  2008\\
		 Mulhouse, France}

\begin{document}

\section{Introduction}

The theory of Quantum Chromodynamics (QCD), underlying the quark and 
gluon structure of hadrons, is invariant under time-reversal (T), which
allows the distinction of T-even and T-odd quantities. Single-spin
asymmetries are specific examples of T-odd observables.  In general,
however, T cannot be used as a constraint for the (complex) S-matrix.
At high energies the cross sections of a (hard) scattering process 
factorizes in
(hadronic) soft parts, which we will discuss in detail, and a partonic
cross section. The partonic cross sections at tree-level are T-even
with T-odd phenomena coming in at order $\alpha_s$ beyond tree-level.
Thus, one expects at leading order T-even soft parts (or an even number
of T-odd parts) to be relevant in double spin asymmetries 
and T-odd soft parts (at least an odd number of them) to be relevant in
single spin asymmetries.

In situations in which one integrates over all transverse momenta one
has collinear correlators describing the soft parts, parametrized in 
terms of distribution 
functions (DF) and fragmentation functions (FF) that depend only on a 
collinear momentum fraction. For spin 0 and spin 1/2 hadrons these 
collinear correlators are only T-even. This leads to the expectation that
T-odd phenomena are absent in high-energy scattering processes.
This indeed is true for processes with {\em only one} hadron in 
the {\em initial} state. 
{\em The} example of this is of course inclusive deep inelastic 
scattering, $ep \rightarrow eX$. 
For hadron-hadron collisions or semi-inclusive
fragmentation processes even a factorized description allows T-odd
phenomena. 

Single spin asymmetries (SSA) are known to exist, persisting at high 
energies~\cite{Adams:1991rw}.
%,Adams:1991cs,Bravar:1996ki,
%Airapetian:2001eg,Adler:2003pb,Adams:2003fx,Airapetian:2004tw}.
There are many studies of mechanisms that lead to 
single spin asymmetries (SSA) in hard scattering processes, with
Sivers~\cite{Sivers:1989cc} and 
%,Sivers:1990fh} and 
Collins effects~\cite{Collins:1992kk,Collins:1993kq} serving
as notable examples in the supposedly {\em simple case} of leptoproduction.
As alluded to, in collinear approximation
(integrating over all transverse momenta) all leading twist distribution
(and fragmentation functions) only depend on the longitudinal momentum
fraction $x$ (or $z$) are T-even. For spin 0 and spin 1/2 hadrons this
implies e.g.\ that
polarized quarks are only found in polarized hadrons (and vice versa). 
Single spin asymmetries (SSA) can occur but restricting ourselves
to collinear correlators require twist-three correlators involving 
quark-gluon matrix elements~\cite{Jaffe:1993xb}.
For higher spins, e.g.\ for spin 1 hadrons, one also has T-odd collinear 
twist-two fragmentation functions~\cite{Bacchetta:2000jk}.
Returning to the (T-odd) quark-gluon matrix elements, one finds that 
they can appear at leading order in the specific limit of a zero-momentum 
gluon, referred to as {\em gluonic pole matrix elements} such as the 
Qiu-Sterman matrix elements~\cite{Qiu:1991pp}.
Also in model calculations the effects of these soft gluon interactions
between the target remnant and the hard part have been
demonstrated, giving rise to distinct effects for initial or final state 
interactions~\cite{Brodsky:2002cx}.

Going beyond the collinear approximation, the incorporation of
intrinsic transverse momenta of partons provides another mechanism to 
generate leading order SSA, which can be traced back to correlations 
between the intrinsic transverse motion and spin of partons and/or 
hadron~\cite{Sivers:1989cc,Collins:1992kk}.
%,Sivers:1990fh,Collins:1992kk}.
The effects are described by transverse momentum dependent 
(TMD) distribution functions~\cite{Mulders:1995dh,Boer:1997nt,Bacchetta:2006tn}, 
containing both T-even and T-odd parts
and depending on longitudinal momentum fraction
$x$ and the transverse momentum $p_{T}$ as appearing in the Sudakov
decomposition $p = x\,P + p_{T}$ (or $p = (1/z)\,P + p_{T}$ for 
fragmentation). The TMD correlators include Wilson lines, which besides
ensuring gauge-invariance are in the case of distribution functions the
sole cause of T-odd contributions. Upon $p_{T}$-integration one finds
after weighing with $p_{T}$ the socalled {\em transverse moments}
of the TMD distribution functions, which can be separated into T-even
and T-odd parts that are universal and of which the T-odd part can be
identified with the gluonic pole matrix elements. 

\section{\label{TMDcorrelators}
Transverse momentum dependent (TMD) correlators}

The TMD distribution functions are projections 
of the TMD quark correlator defined on the light-front
(LF: $\xi{\cdot}n\,{\equiv}\,0$)
\begin{equation}\label{TMDcorrelator}
\Phi_{ij}^{[C]}(x{,}p_\st{;}n)
={\int}\frac{d(\xi{\cdot}P)d^2\xi_\st}{(2\pi)^3}\ e^{ip\cdot\xi}\,
\langle P{,}S|\,\overline\psi_j(0)\,\mathcal U_{[0;\xi]}^{[C]}\,
\psi_i(\xi)\,|P{,}S\rangle\big\rfloor_{\rm{LF}}\ .
\end{equation}
The \emph{Wilson line} or \emph{gauge link}
$\mathcal U_{[\eta;\xi]}^{[C]}\,
{=}\,\mathcal P{\exp}\big[{-}ig{\int_C}\,ds{\cdot}A^a(s)\,t^a\,\big]$
is a path-ordered exponential along the integration path $C$ with
endpoints at $\eta$ and $\xi$, ensuring gauge-invariance.
In the TMD correlator~\ref{TMDcorrelator} the integration path $C$ in
the gauge link turns out to be process-dependent. 

\begin{figure}
\includegraphics[width=3.4cm]{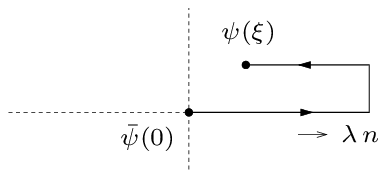}
\hspace{0.5cm}
\includegraphics[width=3.4cm]{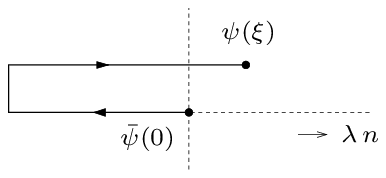}
\\
(a) $\Phi^{[+]}$
\hspace{2.8cm}
(b) $\Phi^{[-]}$
\\[0.2cm]
\includegraphics[width=3.4cm]{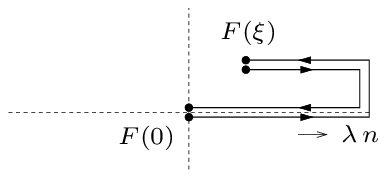}
\hspace{0.3cm}
\includegraphics[width=3.4cm]{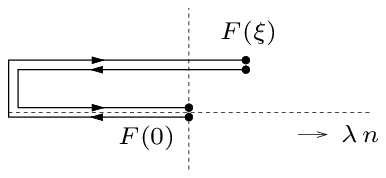}
\hspace{0.3cm}
\includegraphics[width=3.4cm]{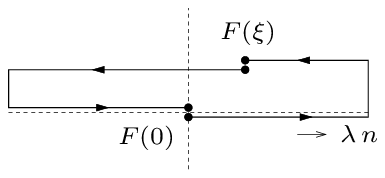}
\hspace{0.3cm}
\includegraphics[width=3.4cm]{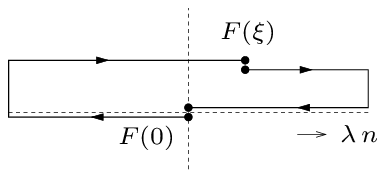}
\\
(c) $\Gamma^{[+,+]}$
\hspace{2.3cm}
(d) $\Gamma^{[-,-]}$
\hspace{2.3cm}
(e) $\Gamma^{[+,-]}$
\hspace{2.3cm}
(f) $\Gamma^{[-,+]}$
%\\[0.2cm]
%\includegraphics[width=3.0cm]{mulders_piet.fig5.eps}
%\hspace{0.5cm}
%\includegraphics[width=3.0cm]{mulders_piet.fig6.eps}
%\\
%(e) $\Gamma^{[+,-]}$
%\hspace{2.1cm}
%(f) $\Gamma^{[-,+]}$
\caption{
Simplest structures (without loops)
for gauge links and operators in quark correlators (a)-(b)
and gluon correlators (c)-(f).
\label{simplelinks}}
\end{figure}
In a diagrammatic approach the Wilson lines arise by resumming all
collinear gluons exchanged between the soft and the hard partonic parts of the
hadronic process.
The integration path $C$ is fixed by the (color-flow structure
of) the hard partonic scattering~\cite{Bomhof:2006dp}.
Basic examples (see Fig.~\ref{simplelinks}) 
are semi-inclusive deep-inelastic scattering (SIDIS) where
for the quark correlator
the resummation of all final-state interactions leads to the future
pointing Wilson line $\mathcal U^{[+]}$,
and Drell-Yan scattering where the initial-state interactions lead to
the past  pointing Wilson line $\mathcal U^{[-]}$.
These links connect the parton fields in the correlator,
running along the light-like direction $n$, conjugate to $P$ 
(satisfying $P\cdot n = 1$ and $n^2 = 0$) and closing 
in the transverse direction at lightcone 
infinity~\cite{Belitsky:2002sm}.
For gluons the correlators including links are 
given by~\cite{Bomhof:2007xt}
\begin{equation}
\Gamma^{[C,C^\prime]}_{\alpha\beta}(x,p_\st;n)
=  \int \frac{d(\xi\cdot P)\,d^2\xi_\st}{(2\pi)^3}
\ e^{i\,p\cdot \xi}
\ \mbox{Tr}\left(F^{n}_{\ \ \beta}(0)\,U^{[n,C]}_{[0,\xi]}
\,F^{n}_{\ \ \alpha}(\xi)\,U^{[n,C^\prime]}_{[\xi,0]}\right)
\vert P\rangle \biggr|_{LF},
\end{equation}
with the simplest possibilities also shown in Fig.~\ref{simplelinks}.

\section{Observables}

Considering intrinsic transverse momenta makes sense because it is
possible to access the effects of them in experiments. 
The collinear fractions ($x$ or $z$) in the 
Sudakov expansion of the parton momenta can be related to kinematical
ratios of hard momenta (e.g. $x \approx x_B = Q^2/2P\cdot q$ 
and $z \approx z_h = P_h\cdot P/P\cdot q$ in semi-inclusive 
deep inelastic scattering) up to $O(1/Q^2)$ corrections.
Therefore the quantity
$q_\st = q + x_B\,P - P_h/z_h \approx k_\st - p_\st$ can be
measured in semi-inclusive deep inelastic scattering (SIDIS), 
$\gamma^\ast (q) + N (P) \rightarrow h (P_h) + X$.
It is zero at leading order ($O(Q)$ in the hard scale), but relates 
to the intrinsic transverse momenta at $O(M)$. Essential in establishing
the relation to the intrinsic transverse momenta is the assumption
of a particular partonic subprocess, in the case of SIDIS 
$\gamma^\ast (q) + q (p) \rightarrow q (k)$. The vector $q_\st$ is the 
transverse momentum of $q$ in a frame in which $P$ and $P_h$ are chosen 
parallel or (experimentally more useful) related to the transverse momentum
of $P_h$, $q_\st = -P_{h\perp}/z_h$ in a frame in which $q$ and $P$ are 
chosen parallel.
With $Q_\st^2 = -q_\st^2$, one needs TMD functions when $Q_\st \sim
O(M)$ and one needs a collinear description involving a subprocess with
one more parton radiated off when $Q_\st \sim O(Q)$. Matching
of these approaches was condidered in Ref.~\cite{Bacchetta:2008xw}.
Not only in electroweak processes like SIDIS or the Drell-Yan process
transverse momenta can be accessed, but one can also consider
inclusive hadron-hadron scattering. 
The experimental signature in this case is the non-collinearity
of the produced particles/jets in the plane perpendicular
to the colliding beam particles, outlined in detail in 
Ref.~\cite{Boer:2003tx}.

Accessing intrinsic transverse momenta in most cases requires a
study of azimuthal dependence in high energy processes. Although
the effects are in principle not suppressed by powers of the hard 
scale in comparison with the leading collinear treatment, it 
requires measuring hadronic scale quantities (transverse momenta)
in a high momentum environment. 
We already mentioned that for the explanation of single spin
asymmetries time reversal invariance plays an important role: 
The T-invariance of QCD allows to distinguish
quantities and observables according to their T-behavior. 
Leading twist collinear correlators $\Phi(x)$ and $\Gamma(x)$ (i.e.\ leading
in an expansion in the inverse hard scale) are all T-even. 
For the TMD correlators, however, the T-operation interchanges
$\Phi^{[+]}(x,p_\st) \leftrightarrow \Phi^{[-]}(x,p_\st)$ (and similar
relations for gluon TMD correlators), allowing us to construct 
correlators with T-even and T-odd operator combinations and 
providing an explanation of SSA.

We note that for fragmentation functions the appearance of an 
hadronic out-state in the definition, prohibits the use of T-symmetry 
as a constraint in the first place. One thus in principle {\em always} had
both T-even and T-odd functions appearing in the parametrization, although 
these do only appear at subleading twist or for TMD correlators.
It is for many purposes still useful to
separate the correlators into two classes containing 
T-even or T-odd operator combinations in analogy with the case 
of distributions, referred to as naive T-even or naive T-odd. But one 
must be aware that both the naive T-even and naive T-odd correlators contain
T-even and T-odd parts and corresponding FFs in the parametrization.

\section{TMD treatment}

As already referred to in section~\ref{TMDcorrelators} the
gauge links in the correlators are the result of resumming
leading matrix elements with collinear gluons. 
The presence of links, differing for each 
partonic sub-diagram and its color-flow,
results in the following generic expression for a hard cross section 
at measured $q_\st$ (involving in general complex diagram-dependent
gauge-link paths),
\be
\frac{d\sigma}{d^2q_\st}\sim \sum_{D,abc\ldots}
\Phi_a^{[C_1(D)]}(x_1,p_{1\st})\,\Phi_b^{[C_2(D)]}(x_2,p_{2\st})
\,\hat\sigma_{ab\rightarrow c\ldots}^{[D]}
\Delta_c^{[C_1^\prime(D)]}(z_1,k_{1\st})\ldots + \ldots
\label{basic}
\ee
where the sum $D$ runs over diagrams distinguishing also the
color flow and $abc\ldots$ is the summation over quark
and antiquark flavors and gluons. Dirac and Lorentz indices,
as well as traces are suppressed. The ellipsis at the end indicate
contributions of other hard processes.

The results for cross sections after integration over the transverse
momenta $q_\st$ also allows integration over the partonic transverse momenta
and one obtains the {\em path-independent} collinear correlators
$\Phi(x)$ rather than
the path-dependent TMD correlators $\Phi^{[C(D)]}(x,p_\st)$. Thus, from
Eq.~\ref{basic} one gets the well-known collinear result
\be
\sigma \sim \sum_{abc\ldots} \Phi_a(x_1)\,\Phi_b(x_2)
\,\hat\sigma_{ab\rightarrow c\ldots}\Delta_c(z_1)\ldots + \ldots,
\ee
where
$\hat \sigma_{ab\rightarrow c\ldots}
= \sum_{D} \hat\sigma^{[D]}_{ab\rightarrow c\ldots}$
is the partonic cross section.

Constructing a weighted cross section (azimuthal asymmetry) by
including a weight $q_\st^\alpha$ in the $q_\st$-integration leads
to 
\be
\langle q_\st^\alpha\,\frac{d\sigma}{d^2q_\st}\rangle
\sim \sum_{D,abc\ldots}
\Phi_{\partial a}^{\alpha [C_1(D)]}(x_1)\,\Phi_b(x_2)
\,\hat\sigma_{ab\rightarrow c\ldots}^{[D]}
\Delta_c(z_1)\ldots + \ldots
\label{basic2}
\ee
containing 
a number of terms in each of which for one of the partons a 
{\em transverse moment} appears,
\be
\Phi_\partial^{\alpha\,[C]}(x) =
\int d^2p_\st\ p_\st^\alpha\Phi^{[C]}(x,p_\st)
=\widetilde\Phi_\partial^{\alpha}(x)
+ C_G^{[U(C)]}\,\pi\Phi_G^\alpha(x,x),
\label{decomposition}
\ee
appears.  These transverse moments still contain a path-dependence, 
so Eq.~\ref{basic2} cannot be simplified immediately but, as also 
shown in the above equation, the path dependence is contained in 
a (gluonic pole) factor $C_G$, which can easily be calculated. The 
first term, $\widetilde\Phi_\partial(x)$, is a collinear correlator
containing matrix elements with T-even operators, while $\Phi_G(x,x-x_1)$ 
is a collinear correlator with a structure of a quark-gluon-quark
correlator involving the gluon field $F^{n\alpha}$.
In Eq.~\ref{decomposition} one needs the zero-momentum ($x_1 = 0$) limit
for the gluon momentum. This matrix element is known as the gluonic pole
matrix element. The operators involved are T-odd. Both collinear
correlators on the RHS in Eq.~\ref{decomposition} are link-independent.
Using this decomposition one can write down a 
parton-model like expansion for the single-weighted cross section
$\left< q_\st^\alpha\sigma\right>$ in Eq.~\ref{basic2} in which 
$\widetilde\Phi_{\partial}^{\alpha}(x)$ is multiplied with the partonic
cross section, while $\pi\Phi_{G}^{\alpha}(x,x)$ is multiplied
with the {\em gluonic pole cross section},
\be
\hat \sigma_{[a]b\rightarrow c\ldots}
= \sum_{D} C_G^{[U(C(D))]}\hat\sigma^{[D]}_{ab\rightarrow c\ldots}, 
\ee
which besides the normal partonic cross sections 
constitutes a different gauge-invariant combination of the squared 
amplitudes~\cite{Bomhof:2006ra}. 
For a number of processes, the consequences for (weighted) azimuthal 
single spin asymmetries have been 
investigated~\cite{Bacchetta:2007sz}.
%,Boer:2007nd,Boer:2007nh,Bomhof:2007su,Lu:2008qu}.
For more complex weightings or
trying to stay at the unintegrated level, one has to make additional
assumptions outlined in Ref.~\cite{Bomhof:2007xt}.
In this paper also the split-up of TMD functions in 
\be
\Phi^{[U]}(x,p_\st) = 
\Phi^{[{\rm even}]}(x,p_\st) + 
G_G^{[U]}\,\Phi^{[{\rm odd}]}(x,p_\st) + 
\delta\Phi^{[U]}(x,p_\st) ,
\ee
with $\Phi^{[{\rm even/odd}]} = \frac{1}{2}(\Phi^{[+]} \pm \Phi^{[-]})$
is discussed. The even and odd combinations are constructed from
the correlators with simple links shown in Fig.~\ref{simplelinks}.
Upon integration one has
$\Phi^{[{\rm even}]}(x) = \Phi(x)$,   
$\Phi^{[{\rm odd}]}(x) = 0$,
$\Phi_\partial^{\alpha\,[{\rm even}]}(x) 
= \widetilde \Phi_\partial^{\alpha}(x)$, and 
$\Phi_\partial^{\alpha\,[{\rm odd}]}(x) = \pi\Phi_{G}^{\alpha}(x,x)$. The
additional terms is referred to as {\em junk TMD} satisfying 
$\delta\Phi^{[U]}(x) =  \delta\Phi_\partial^{\alpha[U]}(x) =  0$.

Similar combinations of even and odd correlators can be constructed
from the simples gluon correlators~\cite{Bomhof:2007xt}, although
one must be aware that there are two types of gluonic pole matrix 
elements~\cite{Ji:1992eu}
corresponding to the two different ways to construct color singlets for
three gluons and correspondingly, there are for instance two distinct
gluon-Sivers distribution functions. 

\section{Conclusions}

The approach to understand T-odd observables like single spin asymmetries
via the TMD correlators and the non-trivial gauge link structure unifies
a number of approaches to understand such observables, in particular
the collinear approach and the inclusion of soft gluon interactions.
Although the treatment of fragmentation correlators also separates into
naive T-even and naive T-odd parts with T-even and T-odd operator structure
respectively, the gluonic pole contributions (naive T-odd parts) 
in the case of fragmentation might very well vanish.
Indications come from the soft-gluon approach~\cite{Collins:2004nx}
and a recent spectral analysis in a spectator model
approach~\cite{Gamberg:2008yt}.

\end{document}